\documentclass[preprint,aps,showpacs]{revtex4}

\usepackage{graphicx}

\usepackage{bm}

\begin{document}

\title{Relativistic particle
incoherent scattering by the nuclei of crystal plane atoms}

\author{Victor V. Tikhomirov}

\affiliation{Research Institute for Nuclear Problems, Belarusian
State University, Minsk, Belarus}

\date{\today}

\begin{abstract}
A consistent theory, which describes the incoherent scattering of
classically moving relativistic particles by the nuclei of crystal
planes without any phenomenological parameter is presented. The
basic notions of quantum mechanics are applied to introduce a
fundamental compact formula for the mean square incoherent
scattering angle per unit length of particle trajectory. The
latter is used to implement the effects of the crystal atom
distribution inhomogeneity into the Coulomb scattering simulations
without noticeable elongation of the simulation time. The theory
essentially reconsiders the nature of positively charged particle
dechanneling from the low nuclear density regions, being essential
in both the crystal undulators and envisaged measurements of the
specific electromagnetic momenta of short living particles.

\end{abstract}

\pacs{61.85.+p,12.20.Ds}

\maketitle

\section{Introduction}

When a fast particle is moving along a string or plane of atoms in
a crystal, nanotube or other ordered structure, it experiences
correlated collisions with successive atoms which result in the
strong particle deflection described by an effective field
\cite{lin} exceeding any created in laboratory by the orders of
value \cite{bar,bar2,akh,bai,kim}. Such strong fields provide a
unique tool for the high energy particle beam manipulations, such
as collimation \cite{tsy,tar2,tik,tik12,bir,sca,tik3}, focusing
\cite{and,sca2,tik14}, extraction \cite{els,lan,maz,syt}; hard
narrow spectrum gamma-radiation, including coherent bremsstrahlung
\cite{ter,akh,bai}, string-of-string \cite{bai2,tik5}, channeling
\cite{kum,bar,akh}, parametric \cite{bar,PXR} and crystal
undulator radiation \cite{bar3,bar,kap,kor,bar4,tik2,bag2,cam};
both the coherent \cite{ter,akh,bai} and synchrotron-like
electron-positron pair production \cite{bar6,bar2,kim,bai};
quantum electrodynamic effects of both electron and positron
magnetic moment modification \cite{bar5,tik13}, vacuum dichroism,
birefringence \cite{bar6,bar,bar2,bai} and some others
\cite{kim,ugg,bai}; production, manipulation and analysis of
polarized gamma \cite{bar,bar6,bar2,bai} and electron/positron
beams \cite{bar7,bar8,bar9,bar14,tik4,tik5}; the reduction of the
thickness of particle detectors and making them sensitive to both
direction and polarization as well as revealing the new factors
which determine their energy resolution \cite{bar10,ban}, search
and measurement of both magnetic and electric dipole as well as
electric quadrupole momenta of short living particles
\cite{bar,bar7,bar11,bag,ner,bar12,fom}, etc.

All the mentioned coherent effects experience an impact of the
incoherent scattering. Though the latter has much in common with
the scattering by isolated atom or atom in amorphous medium,
observable modifications of relativistic particle incoherent
scattering in crystals arise due to both the atom and electron
distribution inhomogeneity in the impact parameter plane,
characterized by the r.m.s. amplitude $u_1$ of atom thermal
vibrations. The latter give rise to the large correlations of the
small momentum transfers $q \leq \hbar/u_1$, which merge together
into a smooth coherent scattering process, described by the
Lindhard potential \cite{lin}, ceasing to contribute to the mean
square angle of incoherent scattering, the decrease of which is
directly measurable.

Though the problem of the specific incoherent scattering in
crystals has been known since 50-th \cite{ter,ter2} and addressed
by several authors \cite{baz,baz2,lju,bar13,tik6,tik7,art}, the
difference of the incoherent scattering in crystals from the
scattering in amorphous medium has been explicitly demonstrated
only recently \cite{maz2}. The experiment has been conducted at
the incidence angles, large enough to simultaneously simplify the
particle motion, scattering theory and reliable extraction of the
7\% root mean square angle reduction in a $30 \mu m$ Si crystals.
However, being obtained as an average over the straight-line
particle trajectories with arbitrary impact parameters for the
whole crystal, the measured value has given only an average
scattering characteristic for a uniform particle flux. The latter,
in general, differs from local scattering characteristics met by
both negatively charged channeling particles inside the dense
nuclear regions and positively charged ones far outside the
latter. The newly available experiments \cite {maz,ban2,syt} in
electron deflection and radiation both need a quantitative theory
of dechanneling of electrins and provide at the same time the
abundant data for its verification. The problem of stable
channeling of positively charged particles is related with the
strong violation of the proportionality of incoherent scattering
intensity to the local nuclear density \cite{lin,mat,tik7},
related to the sharp decrease of the latter, proceeding greatly
ahead of the same of the fields of all the distant plane atoms,
becoming the leading source of incoherent scattering.

To be both correct and practical, the theory of relativistic
particle scattering has to combine the semiclassical scattering
treatment \cite{lin} of particle motion in the averaged potential
with the quantum one \cite{bor,LL3} for its perturbation by a
single atom potential. This problem has been already treated
\cite{tik7} using the Wigner function approach in the axial case.
Though the latter is more general and symmetric, such promising
applications as short living particle electromagnetic moment
measurement, crystal undulator development, channeled electron
beam radiation, extraction and focusing need the incoherent
scattering theory in planar case. Contrary to the axial one, the
latter allows for an explicit treatment of the particle transverse
phase space using the 1D wave functions, making possible a more
reliable approach to the incoherent scattering problem.

Being applied to the incoherent scattering problem, the Wigner
function has to reduce to the phase space density, equal to the
product of a local nuclear density by the differential scattering
cross section, in the classical limit. However, the quantum
evaluation demonstrates negative values, making it necessary to
resort to the positively determined mean square scattering angles
for the small momentum transfer simulations \cite{tik7}. At the
same tine, the large momentum transfers need the modified Coulomb
cross section for their simulation. To avoid introduction of the
corresponding special value, one can combine the high momentum
cross section correction with the low momentum mean square
modification angle into a single additional term to the Coulomb
logarithm.

The paper is organized as follows. The channeling particle wave
function perturbation by the incoherent scattering by a single
atom is introduced in Section II. Along with the Coulomb
scattering basics, quoted in Appendix A, the increase of the
energy of transverse channeling motion, induced by the incoherent
scattering, is used in Section III to extract the local mean
square incoherent scattering angle per unit length, compared with
the 2D theory of \cite{tik7} in Appendix B. The effects of
scattering modification at all the momentum transfers in the
presence of the plane atom distribution inhomogenuity are
represented in the form of a Coulomb scattering logarithm
correction in Section IV. The specific nature of the qualitative
modification of incoherent scattering in the low nuclear density
regions is addressed in both Section IV and Appendix C.

\section{A channeling perturbation by a single atom}

To be both correct and effective, the theory of incoherent
particle scattering in inhomogeneous medium should combine the
unremovable quantum treatment of single atom scattering
\cite{bor,lan} with the classical picture of ultrarelativistic
particle motion in the average potential \cite{lin,bar},
essentially simplifying all the consideration at high energies.
The most reliable way to introduce the latter correctly, is a
grounded reduction of the basic quantum approach. To arrive to the
mean square scattering angle of a classically moving particle, we
will proceed from the average increment of the energy of
transverse particle motion. Being, for instance, confident in the
possibility to neglect the quantum features of transverse
oscillatory channeling motion at high enough energies
\cite{lin,bar}, one can limit consideration to the particle
incoherent scattering by a single atomic plane.

The exhaustive treatment of both the channeling and related
phenomena \cite{bar} proceeds from the "squared" Dirac equation
\cite{LL4} (the system of units $c = \hbar = 1$ is used through
the paper)
\begin {equation} 
\left[ \Delta + p^2 - 2 \varepsilon U(\textbf{r})\right]
\Psi(\textbf{r}) = 0,
\end{equation}
in which $\varepsilon$ and $p$ are particle energy and momentum
respectively, $\Psi(\textbf{r})$ and $U(\textbf{r})$ are particle
wave function and potential energy (potential for short) at a
point $\textbf{r}$. Unperturbed planar channeling is described by
the one-dimensional planar potential $U(x)$, in which the
coordinate $x$ is measured along the plane normal.

Most of the multiple channeling applications involve bent
crystals, in which the effective planar potential becomes
asymmetric and depends on the particle energy. At this, such bent
crystal applications as crystal undulators and measurements of
short living particle specific electromagnetic momenta, reach
their maximal efficiency at the extreme bending, illustrated by
Fig. 1, which makes the dechanneling problem critical even for the
most deeply channeled positively charged particles. Provided the
nuclear density becomes negligible near the potential minima, the
dechanneling theory describing the particle scattering by the
thermal perturbations (singular phonons) of the atoms of the
nearest crystal plane will be developed below mostly in view of
these promising channeling applications.

The solution
\begin {equation} 
\Psi(\textbf{r}) = e^{ip_z z}\varphi(x)
\end{equation}
 of Eq.(1) with the planar potential $U(\textbf{r}) = U(x)$
includes the wave function $\varphi(x)$ of transverse motion,
which obeys the "relativistic Schr$\ddot{o}$dinger equation"
\begin{figure} 
\label{Fig1}
 \begin{center}
 \resizebox{90mm}{!}{\includegraphics{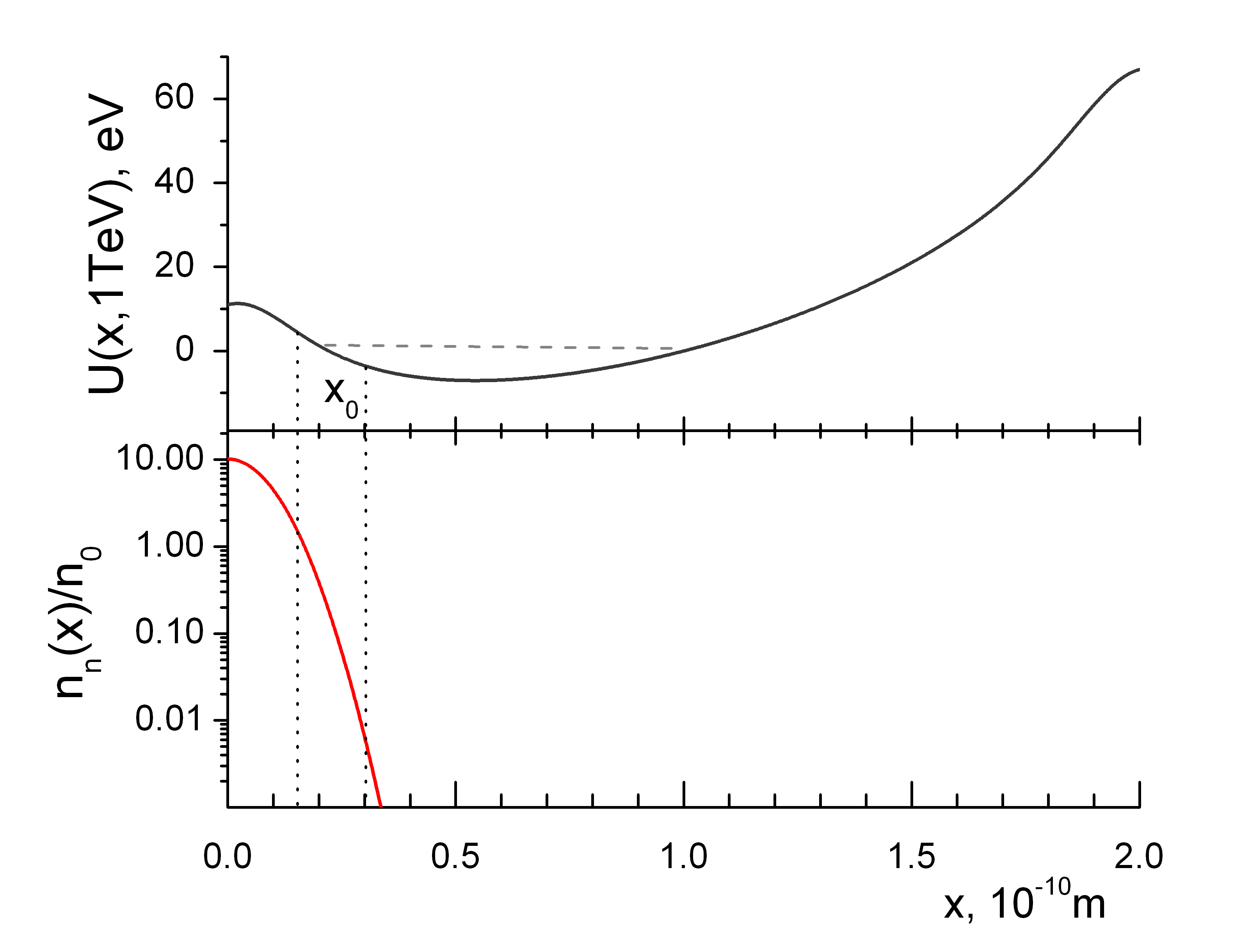}} \\
\caption{The x-coordinate dependence of (110) Ge plane effective
potential at 1 TeV particle energy, bent with 5 m radius (top),
and the averaged planar nuclear density, measured in units of its
average crystal value $n_0$ (bottom). }
\end{center}
\end{figure}
\begin {equation} 
\hat{H}_0\varphi (x) = \left[\frac{\hat{p^2_x}}{2 \varepsilon} +
U(x) \right]\varphi(x) = \left[-\frac{1}{2 \varepsilon}
\frac{d^2}{d x^2} + U(x) \right]\varphi (x) = \varepsilon_n
\varphi(x),
\end{equation}
where $\hat{p} =-i\partial/\partial x$ and  $\varepsilon_n =
(p^2-p_z^2)/2 \varepsilon$ are the momentum operator and energy of
the transverse motion.

Equation $U(x_0) = \varepsilon_n$ determines the classical turning
point coordinate $x_0$, see Fig. 1. Both the channeling stability
and incoherent scattering theory issues reach the maximum urgency
in the region $x_0 = (2\div3) u_1 \sim 0.2$ \AA~of the sharp
decrease of the average nuclear density of an atomic plane
\begin {equation} 
n_n(x_n) = n_0 d\frac{\exp (-x_n^2 u^2_1 /2 )}{\sqrt{2 \pi }u_1},
\end{equation}
where $n_0$ and $d$ are, respectively, the average crystal atomic
density and inter-plane distance. Since the Gaussian (4) changes
much faster than the average potential at $x_n \simeq x_0 \simeq
(2\div3) u_1$, the linear approximation
\begin {equation} 
U(x,) \simeq \varepsilon_n + U^\prime(x_0)(x-x_0)=\varepsilon_n -
E(x_0)(x-x_0),
\end{equation}
where $E(x_0) = -U^\prime(x_0)/e$ is electric field strength at
the turning point, can be readily applied for the latter, reducing
Eq. (1) to
\begin {equation} 
\left[\frac{d^2}{d x^2} + 2 e E(x-x_0) \varepsilon
\right]\varphi(x) = \left[\frac{d^2}{d \xi^2} - (\xi_0-\xi)
\right]\varphi(\xi) = 0,
\end{equation}
\begin {equation} 
\xi = \xi' x,~~~~~~~~~\xi_0 = \xi' x_0,~~~~~~~~~\xi' \equiv
\frac{d \xi}{dx}  = \sqrt[3]{2e|E|\varepsilon} = const
> 0.
\end{equation}
Eq. (6) solution \cite{LL3}
\begin {equation} 
\varphi(x-x_0) = \sqrt{\frac{2 \pi \varepsilon}{\xi'}} Ai(\xi_0
-\xi) = \sqrt{\frac{\varepsilon}{2 \pi \xi'}}
\int_{-\infty}^\infty  \exp [i (\xi_0 -\xi) t + i t^3/3] dt
\end{equation}
contains an Airy function $Ai$, which decreases sharply in the
under-barrier $x < x_0$ and oscillates fast (see Fig. 18 in
\cite{bar2}) in the classical motion region $x > x_0$ at high
particle energies. The solution (8) has been normalized according
to the condition
\begin {equation} 
\frac{dP(x)}{dx} = \overline{|\varphi(x-x_0)|^2}= v^{-1}(x)
\end{equation}
for the averaged transverse probability density, where
\begin {equation} 
v(x) \simeq  \sqrt{2e|E|(x-x_0)/\varepsilon} = \sqrt{(x-x_0)}\,
\xi^{3/2}/\varepsilon
\end{equation}
is the transverse particle velocity in the linearized potential
(5). Such a normalization will simplify the introduction of the
quantum scattering characteristics for a classically moving
particle below.

An average value of the transverse energy increment, induced by
the particle incoherent scattering, is readily evaluated through
the Schr$\ddot{o}$dinger equation of the transverse particle
motion
\begin {equation} 
i \frac{d\Psi}{dx} = \hat{H}\Psi,~~~~~~~~~\hat{H} = \hat{H}_0 +
\delta U(x-x_n,y-y_n,z-z_n),
\end{equation}
perturbed by the residual atomic potential \cite{baz,baz2}
\begin {equation} 
\delta U(\textbf{r} - \textbf{r}_n) = U_{at}(\textbf{r} -
\textbf{r}_n) - \int \int U_{at}(\textbf{r} - \textbf{r}_n)
n_n(x_n,y_n,z) dx_n dy_n,
\end{equation}
where
\begin {equation} 
U_{at}(\textbf{r} - \textbf{r}_n) = \frac{Z\alpha}{|\textbf{r} -
\textbf{r}_n|}e^{-|\textbf{r} - \textbf{r}_n| \kappa}
\end{equation}
is the screened atomic potential with the screening constant
$\kappa = 1/r_{sc}$ and radius $r_{sc}$ and
\begin {equation} 
\displaystyle  { n_n(x_n,y_n,z) = \frac{1}{2 \pi u^2_1 d_{at} }
\exp \left\{-x_n^2 u^2_1/2 - [y_n \cos \psi + (z-z_{ax})\sin
\psi]^2  u^2_1/2 \right\}}
\end{equation}
the atomic string nuclear density, characterized by both the atom
thermal vibration amplitude $u_1$ and interatomic distance
$d_{at}$, while $z_{ax}$ is the coordinate of the string
intersection with the plane $xz$ of channeling oscillations.
Though we consider the planar channeling, the string atomic
structure inside a plane (see Fig. 2) should be taken into
consideration to treat the incoherent scattering in both $xz$ and
$yz$ planes adequately.

\begin{figure} 
\label{Fig2}
 \begin{center}
\resizebox{40mm}{!}{\includegraphics{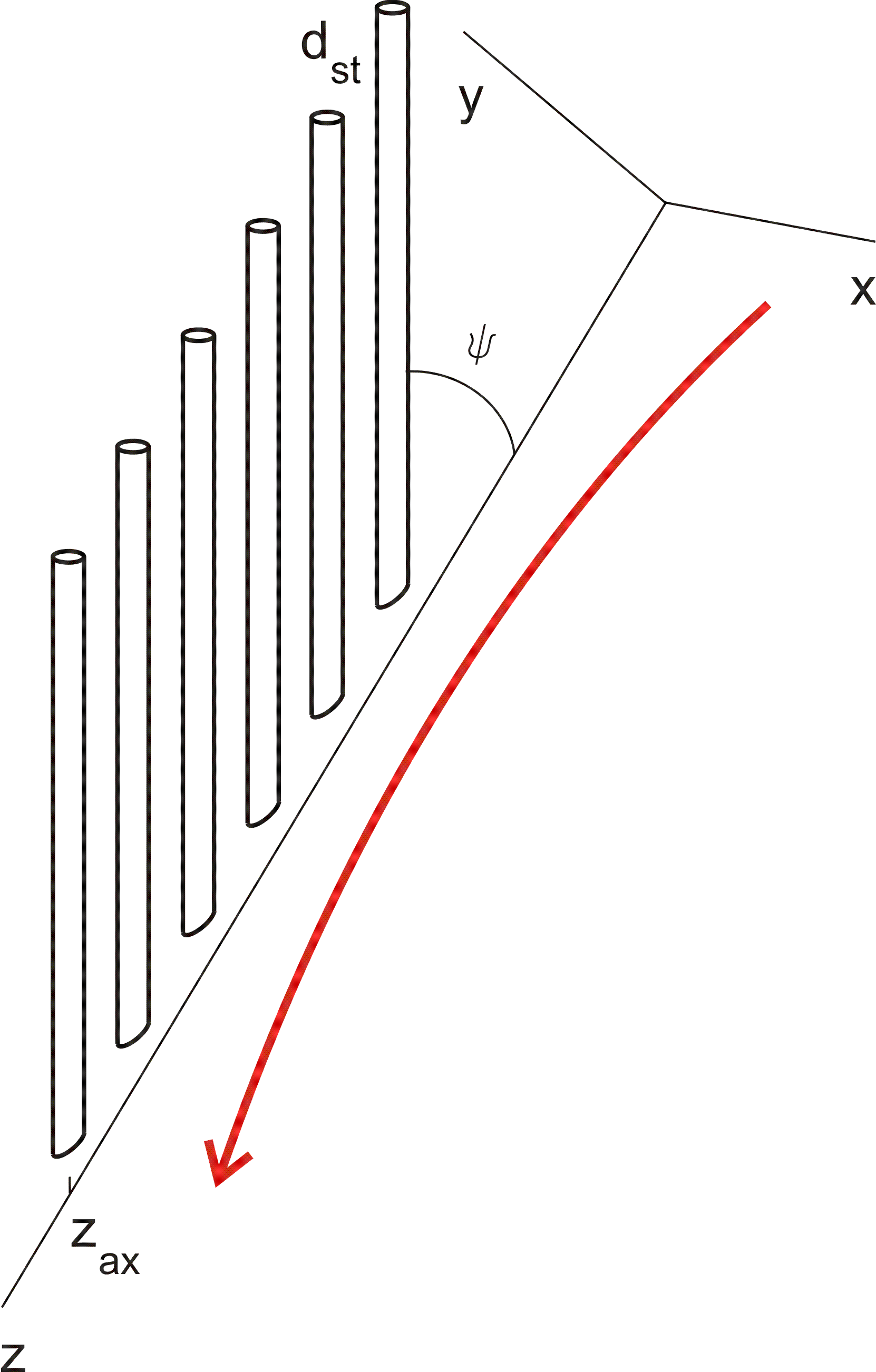}} \\
\caption{Particle motion in the field of atomic strings
(cylinders), constituting an atomic plane, parallel to $yz$
coordinate plane. Planar channeling motion (thick red curve)
occurs in $xz$ plane, constituting a small angle $\psi \ll 1$ with
the atomic strings, which form $yz$ plane.}
\end{center}
\end{figure}

A substitution $ \Psi(\textbf{r}) = e^{i p_z z} f(\textbf{r})
\varphi(x)$ introduces a wave function modification factor
$f(\textbf{r})$ which obeys the equation
\begin {equation} 
i \frac{\partial f(\textbf{r})}{\partial z} =
\frac{\varepsilon}{p_z} \delta U(\textbf{r}  - \textbf{r}_n)
f(\textbf{r}) -\frac{1}{p_z}\frac{\partial f(\textbf{r})}{\partial
x} \frac{d \varphi(x)}{d x} - \frac{1}{2 p_z}\Delta_{\bot}
 f(\textbf{r}) \simeq \frac{1}{v}
\delta U(\textbf{r} - \textbf{r}_n)  f(\textbf{r}).
\end{equation}
In general, both the full Eq. (15) and factor $f(\textbf{r})$
describe both channeling and above-barrier motion of the particles
scattered by an atom at various angles in a complex way. However
to treat the average transverse energy increase, one can consider
only the right hand side limit of Eq. (15), resulting in the
solution
\begin {equation} 
f(x-x_n,y-y_n,z=z_n +0) \simeq \exp \left( -i
\int_{-\infty}^\infty \delta U(x-x_n,y-y_n,z) \frac{dz}{v}
\right),
\end{equation}
applicable just behind the scattering atom, where the complex
evolution of the particle flux distribution has no distance to
develop and only the momentum instant change, described by the
same of the wave function phase
\begin {equation} 
\begin{array}{c}
\displaystyle  {\int_{at}\delta U(x - x_n, y - y_n, z - z_n)
\frac{dz}{v} }
\\\displaystyle  { = \frac{Z \alpha}{\pi v} \int \int
\left (e^{ -i \chi_x x_n - i \chi_y y_n} - e^{ - (\chi_x^2 +
\chi_y^2 \sec^2 \psi )u_1^2/2 -i k_y z \tan \psi }\sec \psi \right
) e^{ i \chi_x x + i \chi_y y } \frac{d\chi_x d\chi_y}{\chi^2 +
\kappa^2} }
\end{array}
\end{equation}
should be considered. The found solution $ \Psi(\textbf{r}) = e^{i
p_z z} f(\textbf{r}) \varphi(x)$ will make it possible to evaluate
the average increment of channeling particle transverse energy in
Section III.

Before to evaluate the latter, a few points, related with the
atomic string consideration, should be clarified. First, the
interatomic $d_{at}$, inter-string $d_{st}$ and inter-plane $d$
distances satisfy altogether the condition $n_0 = d_{at} d_{st}
d$. The planar nuclear density (4) is related with the axial one
by the averaging over the distance $d_{st}/\sin \psi $, passed by
a particle between the neighbouring atomic string:
\begin {equation} 
\displaystyle  { n_n(x_n) =  \int_{-\infty}^\infty
n_n(x_n,y_n,z-z_{ax})  \frac{dz }{d_{at}/\sin \psi}. }
\end{equation}
In general, the choice of the string orientation angle $\psi$ from
Fig. 2 can considerably influence the planar channeling. Namely,
at $\psi \leq \vartheta^{ax}_{ch}$, where $\vartheta^{ax}_{ch}$ is
the angle of axial channeling, the latter takes place, while at
$\psi \gg 1^\circ$ the irregular scattering by the high-index
strings of the plane $yz$ becomes possible \cite{maz2}. To make
the planar channeling certainly regular, we assume below that
$\vartheta^{ax}_{ch} \ll \psi \leq 1^\circ$, assuring both the
absence of the above-mentioned effects and the easily predictable
picture of the incoherent scattering in the $yz$ plane, also
discribed by Eq. (17). Finally, the inequality $1 - \cos \psi \ll
10^{-3}$ makes it possible to put both $\sin \psi = \psi$ and
$\cos \psi =1$ below.

\section{Mean square angle of the incoherent scattering of a classically moving particle}

The average transverse energy increment, induced by the incoherent
particle scattering by nuclei, is applied to introduce the mean
square scattering angle per a unit length of a classically moving
particle trajectory in this section. Both the solutions of Eqs.
(6), (15) and the residual atomic potential property
\begin {equation} 
\int \int \delta U(\textbf{r} - \textbf{r}_n) n_n(x_n,y_n,z) dx_n
dy_n = 0
\end{equation}
allow one to represented the average incoherent transverse energy
increment over a unit time, corresponding to the crystal length
$\Delta z = v$, in the form
\begin {equation} 
\begin{array}{c}
\displaystyle  { \langle \Delta \varepsilon_n \rangle = \int_0^v
\int_{-\infty}^\infty \int_{-\infty}^\infty \left (\int \psi^*
\hat{H} \psi \: dx - \int \varphi^*
\hat{H}_0 \varphi \: dx \right) n_n(x_n,y_n,z) dx_n dy_n dz}\\
\displaystyle  { = \frac{\pi v}{\xi^\prime d_{st}}
\int_{-\infty}^\infty \int_{-\infty}^\infty \int_{-\infty}^\infty
\int_{-\infty}^\infty \left(\int_{-\infty}^\infty
\frac{d}{dx}\delta U(x-x_n,y-y_n,z-z_n) \frac{dz}{v} \right)^2
n_n(\rho_n) dx_n dy_n } \\ \displaystyle  {\times{ Ai^2(x-x_0)dx
dy } }.
\end{array}
\end{equation}
The latter demonstrates that the residual potential (12)
introduction is justifies by the considerable Eq. (20)
simplification through the nullification of the terms, linear in
both the first and the second derivatives of the former. The
explicit form of both Eq. (8) and (17) reveals a pair of readily
integrable Dirac functions. Along with the Gauss integration, the
former allows one to reduce Eq. (20) to the form
\begin {equation} 
\begin{array}{c}
\displaystyle  { \langle  \Delta \varepsilon_n\rangle = -\frac{Z^2
\alpha^2 n_0 d}{\pi v^2 \xi'^2} } \int_{-\infty}^\infty
\int_{-\infty}^\infty \int_{-\infty}^\infty  \int_{\frac{|\zeta -
\eta|}{2}< q_2} \frac{e^{-(\zeta + \eta)^2 u^2_1/2} -
 e^{-(\zeta^2 + \eta^2)u^2_1/2}}{(\zeta^2 + \chi_y^2 + \kappa^2) ( \eta^2 +
\chi_y^2 + \kappa^2)}\\
\\
\times  exp \left[i x_0 (\zeta + \eta) + \frac{2i}{3}
 \left( \frac{\zeta + \eta}{2 \xi'}  \right)^3 +i \frac{\zeta + \eta}{\xi'}t^2
  \right ] \zeta d \zeta\: \eta d\eta \: dt \:d\chi_y
\end{array}
\end{equation}
with the common for the Coulomb scattering integration limit,
discussed in Section IV. A novel cubic term of the last Eq. (21)
exponent is related with the quantum under-barrier particle
penetration, being negligible under $\xi' u_1 \gg 1$, or at the
particle energies
\begin {equation} 
\varepsilon \gg \frac{1}{eEu_1^3} < 1 GeV,
\end{equation}
and will be further neglected.

Being an average over the particle positions, the energy increment
(21) is appropriate to estimate both the averaged mean square
scattering angle and dechanneing length. However, to characterize
the incoherent scattering at any point of particle trajectory, one
needs a local value of the mean square scattering angle per unit
length. The latter can be extracted from Eq. (21) using the large
momentum limit (A4) of the mean square angle of Coulomb scattering
at the local nuclear density $n(x)$. To apply the latter, let us
represent Eq. (21) in the form of the integral over the particle
coordinate $x$, readily introduced using the integral
representations
\begin {equation} 
\begin{array}{c}
\displaystyle  {\int_{-\infty}^\infty \exp \frac{2 i k t^2 }{\xi'}
dt = \sqrt{\xi'} \int_{x_0}^\infty \exp \left[2 i k (x-x_0)\right]
\frac{dx}{\sqrt{x-x_0} } } \\ \simeq \displaystyle
{\frac{\xi'^2}{\varepsilon}\int_{x_0}^\infty \exp \left[2 i k
(x-x_0)\right] \frac{dx}{v_x(x)} }.
\end{array}
\end{equation}

Treating the combinations
\begin {equation} 
q_x = \frac{\zeta - \eta}{2},~~~~~q_y =\chi_y,
\end{equation}
\begin {equation} 
q_1 < \sqrt{x_x^2 + q_y^2} < q_2
\end{equation}
as the components of the transferred momentum and representing Eq.
(21) in the form
\begin {equation} 
\displaystyle  { \frac{2 \varepsilon \langle  \Delta
\varepsilon_n\rangle}{p^2} = \int_{x_0}^\infty \left\langle
\frac{d \theta^2_x (x)}{d z}\right\rangle \frac{v dx}{v_x(x)} },
\end{equation}
one can introduce the local mean square projected angle of
incoherent scattering per unit length
\begin {equation} 
\begin{array}{c}
\displaystyle  {\left\langle  \frac{d \theta^2_x (x,q_2,q_1)}{d
z}\right\rangle = \frac{4 Z^2 \alpha^2 n_0 d}{\pi p^2 v^2 }
\int_{q_1}^{q_2} \int_0^{2 \pi} \int_{-\infty}^\infty \frac{\exp
(-2 k^2 u^2_1) - \exp [-(q^2 + k^2)u^2_1]}{[(q_x + k)^2 + q_y^2 +
\kappa^2] [ (q_x - k)^2 + q_y^2 + \kappa^2]} }
\\  \times (q^2_x -k^2) \exp (2 i k x)
dk \: d\varphi \: q dq .
\end{array}
\end{equation}
Besides the integration over the transferred momentum $(q_x = q\:
cos \varphi, q_y=q\:sin \varphi)$, the latter contains the same
over the variable $k=(\zeta+\eta)/2$, taking into consideration
the interference effects in scattering by inhomogeneously
distributed atoms of the crystal plane.

It is shown in Appendix B that Eq. (27) alternatively follows from
the Wigner function approach \cite{tik7}, developed for the axial
case. Both the latter and the present approach make it possible to
get the similar expression for the mean square angle (B4) in the
plane $yz$, opening up alternative methods of the developed theory
experimental verification. The agreement of Eqs. (27) and (B2)
demonstrates both their independence on the particle distribution
and the possibility of the classical treatment of the latter at
high energies.

The lower integration limit $q_1$ has been timely introduced just
to demonstrate that both Eq. (27) and (B4) reduce to the standard
Coulomb scattering theory predictions for the mean square
projected scattering angle (A4) in the high transferred momentum
limit of $q \gg 1/u_1$. Below we put $q_1 = 0$  and concentrate on
the choice of the upper integration limit $q_2$ in Eq. (27).

Using the Feynman integral \cite{LL4} and introducing the
integration variables $q_x \pm k$, Eq. (27) is readily represented
in the form
\begin {equation} 
\begin{array}{c}
\displaystyle {\left\langle  \frac{ \theta^2_x (x, q_2)}{d
z}\right\rangle = \frac{4 \pi Z^2 \alpha^2 n(x)}{p^2 v^2} \left[
\delta \! \ln (x,q_2) + \ln {\left( 1 +
\frac{q^2_2}{\kappa^2}\right)^{1/2}} -
\frac{q^2_2}{2(q^2_2+\kappa^2)}\right] }\\
\displaystyle { =  \frac{2 Z^2 \alpha^2 n_0 d}{p^2 v^2 } \Biggl\{
\int_{-\infty}^\infty \Biggl[\frac{2 k^2-q^2_2}{2 k \sqrt{q^2_2 +
k^2+ \kappa^2} } \ln {\left( \frac{\sqrt{q^2_2 +
 k^2+ \kappa^2} + k}{\sqrt{q^2_2 +  k^2 + \kappa^2} - k} \right) }\; }
\\
\displaystyle  {\;-\frac{k}{\sqrt{k^2+ \kappa^2} } \ln {\left(
\frac{\sqrt{k^2+ \kappa^2} + k}{\sqrt{k^2 + \kappa^2} - k}
\right)\Biggr]
} \exp(2 i k x -2 k^2 u_1^2) dk }\\
\displaystyle { - \frac{2}{\pi} \int_0^{q_2}\left(
 \int_{-\infty}^\infty \frac{ sin (k
x ) \exp(-k^2 u_1^2/2) }{k^2+q^2+\kappa^2 } k dk  \right)^2
\exp(-q^2  u_1^2) d q} \\
 \displaystyle  { + \frac{\sqrt{\pi}}{\sqrt{2 } u_1}  \exp
\left( -\frac{x^2}{2 u^2_1} \right) \ln {\left( 1 +
\frac{q^2_2}{\kappa^2}\right)} \Biggr\} }
\end{array}
\end{equation}
with the integration orders reduced from three to one and two. Eq.
(28) reflects that, following the basic papers
\cite{mol,bet,ter,fan}, all the modifications of the incoherent
scattering process by the atom distribution inhomogeneity can be
grouped into the correction $\delta \! \ln (x,q_2)$ to the
logarithm of the basic Coulomb scattering formulae from Appendix
A.  Fig. 3 demonstrates that $\delta \! \ln (x,q_2)$ gathers its
value mainly at the momentum values $q \sim 1/u_1$, related by the
uncertainty principle with the thermal vibration amplitude $u_1$.
Fig. 3 also demonstrates that $\delta ln(x,q_2)$ becomes negative
at any $q_2$ in the region $x < 2 u_1$, describing the effect of
incoherent scattering reduction there, as well as at $q_2 \leq
1/u_1$ in the region $ x \gg u_1$, demonstrating the need in a
special simulation approach.
\begin{figure} 
\label{Fig3}
 \begin{center}
\resizebox{100mm}{!}{\includegraphics{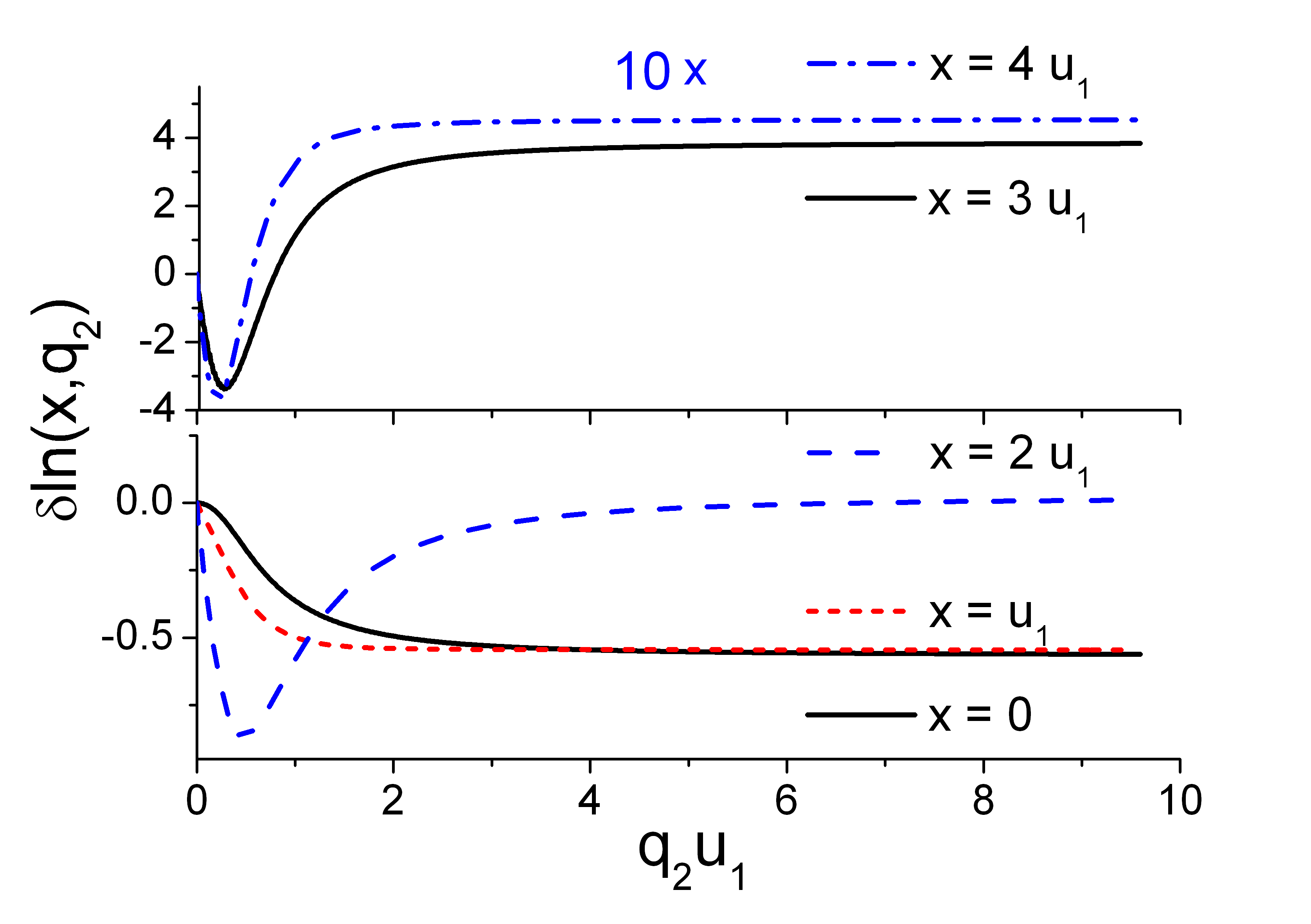}} \\
\caption{Dependence of the Coulomb scattering logarithm
modification on the integration limit of Eq. (28) at $x=0,~ 1,~
2,~ 3$ and $4~u_1$. The case of (110) Si planes at T=293 K is
considered}
\end{center}
\end{figure}

\section{A concise incoherent scattering simulation method}

\subsection{The multiple scattering simulation limited role}

The multiple scattering angle equation (27) will now be
incorporated into the simulation procedure of incoherent
scattering at arbitrary momentum transfers. The former, in fact,
does not provide an adequate simulation tool by itself, though can
be applied for the approach \cite{tar} refinement. It should be
emphasized \cite{tik7}, that the wide applicability of the
multiple Coulomb scattering theory for macroscopic targets in no
way justifies the same for channeling. The point is that the
former is adequate only when the root mean square scattering angle
(A2) exceeds all or most of the single scattering angles, making
the particle deflection looking as a smooth continuous
diffusion-like process \cite{jac,sig}. However the channeling
simulations rely on the thousand and million times shorter
sub-micron trajectory steps within which single scattering angles
mostly far exceed that of the multiple scattering, making the
trajectory steps looking like an angle. This, effectively, "large"
angle single scattering, in particular, is responsible for the
stochastic nature of both the particle trajectories and transverse
energy evolution, including both sudden dechanneling and
rechanneling occasions. None of the latter can be described by the
established multiple scattering theory, making impossible to
introduce the upper limit to Eqs. (A2)-(A4) consistently
\cite{bir}. That is why the multiple scattering simulations must
be reduced to the minimum, determined by the quantum effects
\cite{tik7}.

\subsection{The enhanced power of the Coulomb scattering logarithm modification}

No problems of the cross section definition surely appear in the
multiple Coulomb theory in the uniform medium. However, the
situation changes in the presence of a plane or a string atom
inhomogeneity. Namely, the cross section, formally extracted from
Eqs. (B2)-(B4) following a requirement of its approaching the
Coulomb value (A1) at $q \gg 1/u_1$, proves to be negative at $q
\leq 1/u_1$ \cite{tik7}. This uncommon circumstance makes the
routine simulations impossible and should be considered as a
natural consequence of the transversely "nonlocal" quantum
scattering process reduction to the classical trajectory. However,
since the extracgted cross section attains negative values at the
small momenta $q \leq 1/u_1$ only, they can be readily "absorbed"
by the mean square angle integrals, which can be readily made
positive by the integration limit $q_2$ extension.

The latter, at the same time, secures a positive meaning of the
extracted modified cross section \cite{tik7}, making it valid for
the single scattering sampling at $q > q_2$. However, since the
latter requires additional operations, more effective will be to
sample the unperturbed Coulomb scattering, including the large
momentum scattering modification effect into the mean square
multiple scattering angle. To introduce the latter, the multiple
Coulomb scattering theory prediction (A3) for a uniform medium
with nuclear density $n(x)$ has been separated out in Eq. (28).

As Fig. 3 demonstrates, the logarithm modification value $\delta
ln(x,q_2)$, introduced in Eq. (28) to put together all the effects
of the atom distribution inhomogeneity, converges fast to the
limit
\begin {equation} 
\begin{array}{c}
\displaystyle {\delta \! \ln(x)= \delta ln(x,q_2\gg 1/u_1) =
\frac{u^2_1 - x^2}{3 u^4_1(q^2_2+\kappa^2)} }
\\
\displaystyle { -\Biggl\{ \int_{-\infty}^\infty \frac{k}{
\sqrt{k^2+ \kappa^2} } \ln {\left( \frac{\sqrt{k^2+ \kappa^2} +
k}{\sqrt{k^2 + \kappa^2} - k} \right) } \exp(2 i k x -2 k^2 u_1^2)
dk }
\\
\displaystyle { + \frac{2}{\pi} \int_0^\infty \left(
 \int_{-\infty}^\infty \frac{ \sin (k
x )}{(k^2+q^2+\kappa^2)} \exp \left( \frac{-k^2 u^2_1}{2} \right)
kdk  \right)^2 \exp(-q^2 u_1^2) d q \Biggr\} \frac{u_1 }{\sqrt{2
\pi}} \exp \left( \frac{x^2}{2 u^2_1} \right), }
\end{array}
\end{equation}
reducing Eq. (28) to
\begin {equation} 
\left\langle  \frac{ \theta^2_x (x, q_2)}{d z}\right\rangle =
\frac{4 \pi Z^2 \alpha^2 n(x)}{p^2 v^2} \left[ \delta \! \ln (x) +
\ln {\left( 1 + \frac{q^2_2}{\kappa^2}\right)^{1/2}} -
\frac{q^2_2}{2(q^2_2+\kappa^2)}\right].
\end{equation}
The first term of Eq. (29) is, strictly speaking, negligible at
$q_2 =\vartheta_{max}/p$, where $\vartheta_{max}$ is the maximal
nuclear scattering angle of the logarithmic approximation, and is
shown only to recognize that the preservation of the $\kappa^2$
term in the sum $q^2_2 + \kappa^2$ exceeds the real precision,
serving only the purpose of the Eq. (30) correspondence to the
canonical equation (A3). Eq. (30) makes it possible to locally
describe the incoherent scattering, on the one hand, of the
electrons, moving in the dense nuclear regions, and, on the other,
of both protons and positrons, stably channeling in the negligible
density ones. The average value of the same can be applied to the
case of the uniform particle flux distribution realized at large
incidence angles, described by the Eq. (30) average
\begin {equation} 
\displaystyle  { \overline{\delta \! \ln (x)} \equiv \frac{1}{d}
\int_0^d \delta \! \ln (x) dx = -\int_0^\infty \frac{q^2 \exp
(-q^2 u_1^2)}{2\, (q^2 + \kappa^2)^2} d q^2 \simeq -0.43 } ,
\end{equation}
quantifying the effect of incoherent scattering reduction in
crystals, predicted in \cite{ter2} and measured in \cite{maz2}.

\begin{figure} 
\label{Fig4}
 \begin{center}
\resizebox{90mm}{!}{\includegraphics{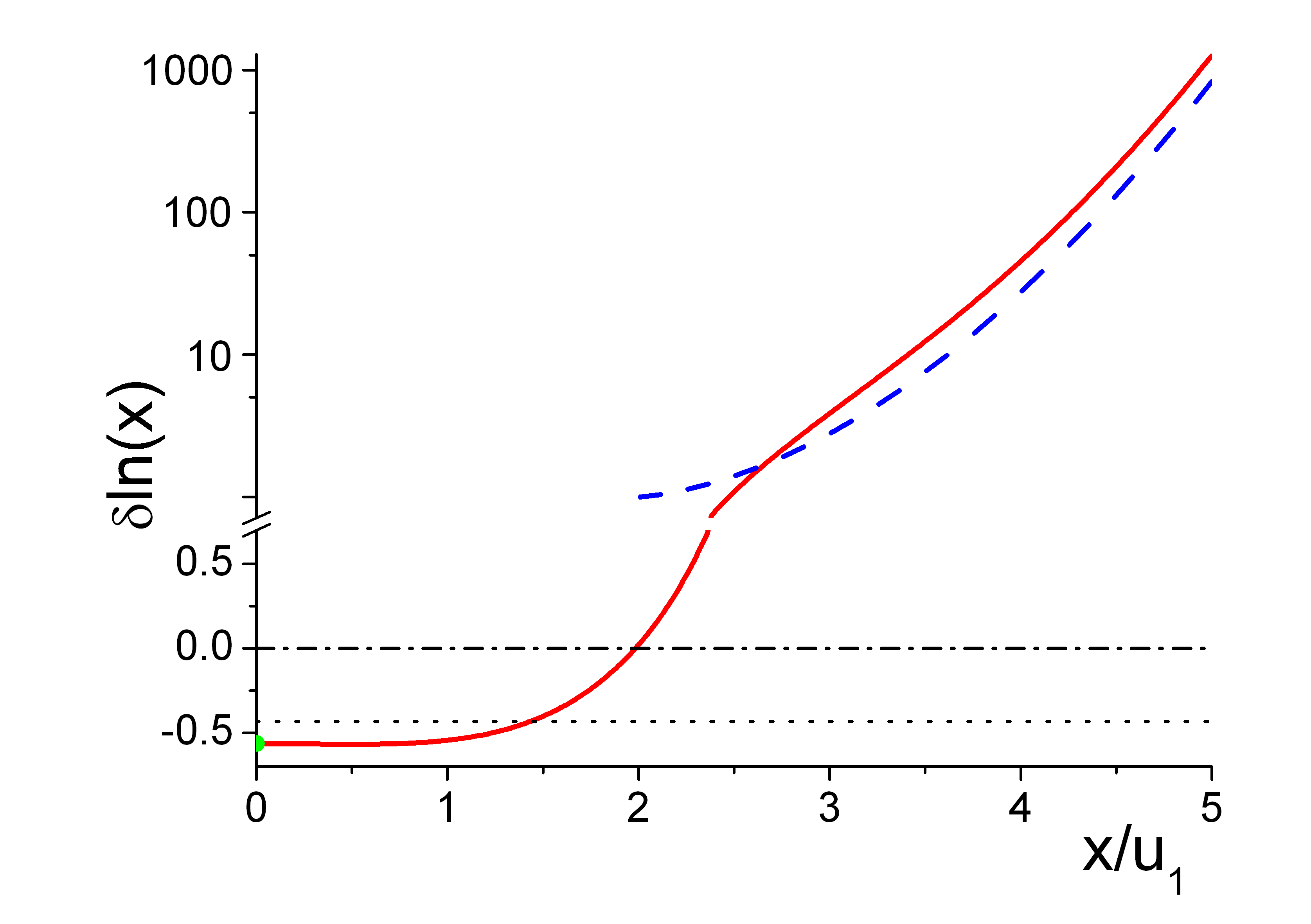}} \\
\caption{ The x-coordinate dependence of the logarithm
modification (29), describing the incoherent scattering difference
from that in amorphous medium of the same density. Dashed line
depicts the asymptote (C5), while dotted one the average value of
the same effect (31) in the high incidence angle limit
\cite{ter2}. The case of (110) Si planes at T293 K is considered.}
\end{center}
\end{figure}

\subsection{The local scattering approximation and its violation at $x \gg u_1$}

Fig. 4 illustrates the coordinate dependence of both the logarithm
modification (29) and its leading expansion term evaluated in
Appendix C, evaluated for the (110) Si plane at $u_1 = 0.075$ \AA,
which demonstrates qualitatively different behavior in the regions
of high, $x < u_1$, and low, $x > 2 u_1$, nuclear density. Indeed,
at $x \leq u_1$ the Gaussian integrals in Eq. (30) form at small
$k \sim 1/u_1$, making it possible to treat (A5) independently
from the integration over $q$ starting from a rather low Eq. (A4)
limit $q_1 \simeq 1/u_1$, revealing the proportionality to the
local nuclear density $n(x)$. The only noticeable modification of
the (27) integral occurs at $q \leq q_1 \simeq 1/u_1$, which,
given the $\kappa^2$ presence in the denominator, results in the
logarithm reduction of $\delta \! \ln (x \leq u_1) \simeq -0.57$.
The latter locally describes the effect of incoherent scattering
reduction, being both closely related to the experimentally
observed \cite{maz2} average effect (31) and directly applicable
to the electron channeling.

Being both natural and indirectly confirmed by the experiment at
$x \leq u_1$, the local scattering approximation causes reasonable
doubts at $x \gg u_1$. Indeed, first, the rapid increase of the
Coulomb denominator at $k \geq \kappa$ disturbs the Gaussian
integral formation, which occurs at the larger $k \sim x/u_1^2 \gg
1/u_1$. Second, the averaged atomic field decreases like the
exponential function $e^{-\kappa x}$, or much slower than the
Gaussian (4). By this reason the thermal atom field fluctuations
must both give the dominant contribution to the local mean square
incoherent scattering angle at $x \gg u_1$ and be proportional to
the product $ u_1^2 e^{-2 \kappa x}$. This mechanism, called
\cite{kit} the one-phonon excitation, was suggested by J. Lindhard
\cite{lin} and verified experimentally \cite{mat,nit,nit2} in the
axial case in 80-th. Fig. 4 illustrates the similar predictions,
obtained in the planar case by the direct numerical evaluation and
confirmed by the analytical estimate in Appendix C.

The revealed incoherent scattering mechanism in no way can be
coordinated with the local approximation in which the local
nuclear density is directly substituted into Eq. (A2). Indeed,
since the formula (A2) embraces all the possible scattering
angles, its solitary use assumes the full neglect of the large
angle catastrophic scattering processes as well as unnatural
reduction of all the stochastic features of particle motion. At
the same tine, despite the drastic Coulomb logarithm
overestimation, the vanishing nuclear density factor results, in
total, in the underestimation of the mean square incoherent
scattering angle at $x > 3 u_1$.

\subsection{The incoherent scattering sampling}

Eqs. (29), (30) make it possible to describe the incoherent
scattering in the regions of both the high and negligible nuclear
density. However, according to the opposite logarithm (29) signs
(see Fig. 4), the simulation algorithms considerably differ at $x
\leq u_1$ and $x \gg u_1$. Indeed, if $\delta \ln(x) < 0 $, the
incoherent scattering process can be sampled in full as a Coulomb
scattering by the angles $q > q_2^0/p$, limited from below by the
momentum transfer $q_2^0$, determined by the equation
\begin {equation} 
\left\langle  \frac{d \theta^2_x (x, q_2^0)}{d z}\right\rangle =
0.
\end{equation}
At this, all the incoherent scattering modifications are taken
into consideration by abandoning of the momentum transfers of $q <
q_2^0$.

\begin{figure} 
\label{Fig5}
 \begin{center}
\resizebox{70mm}{!}{\includegraphics{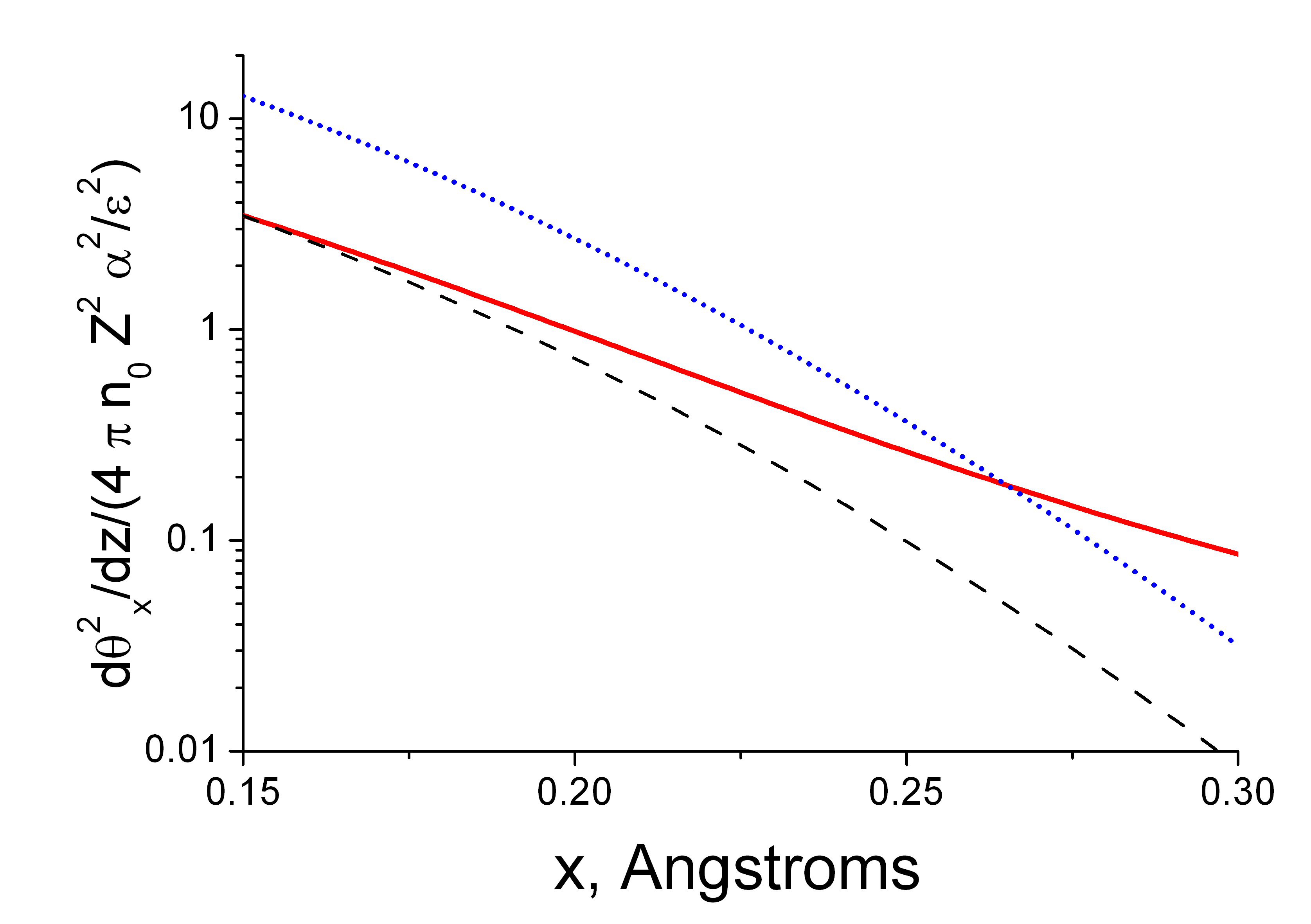}}\hspace{1cm}
\resizebox{70mm}{!}{\includegraphics{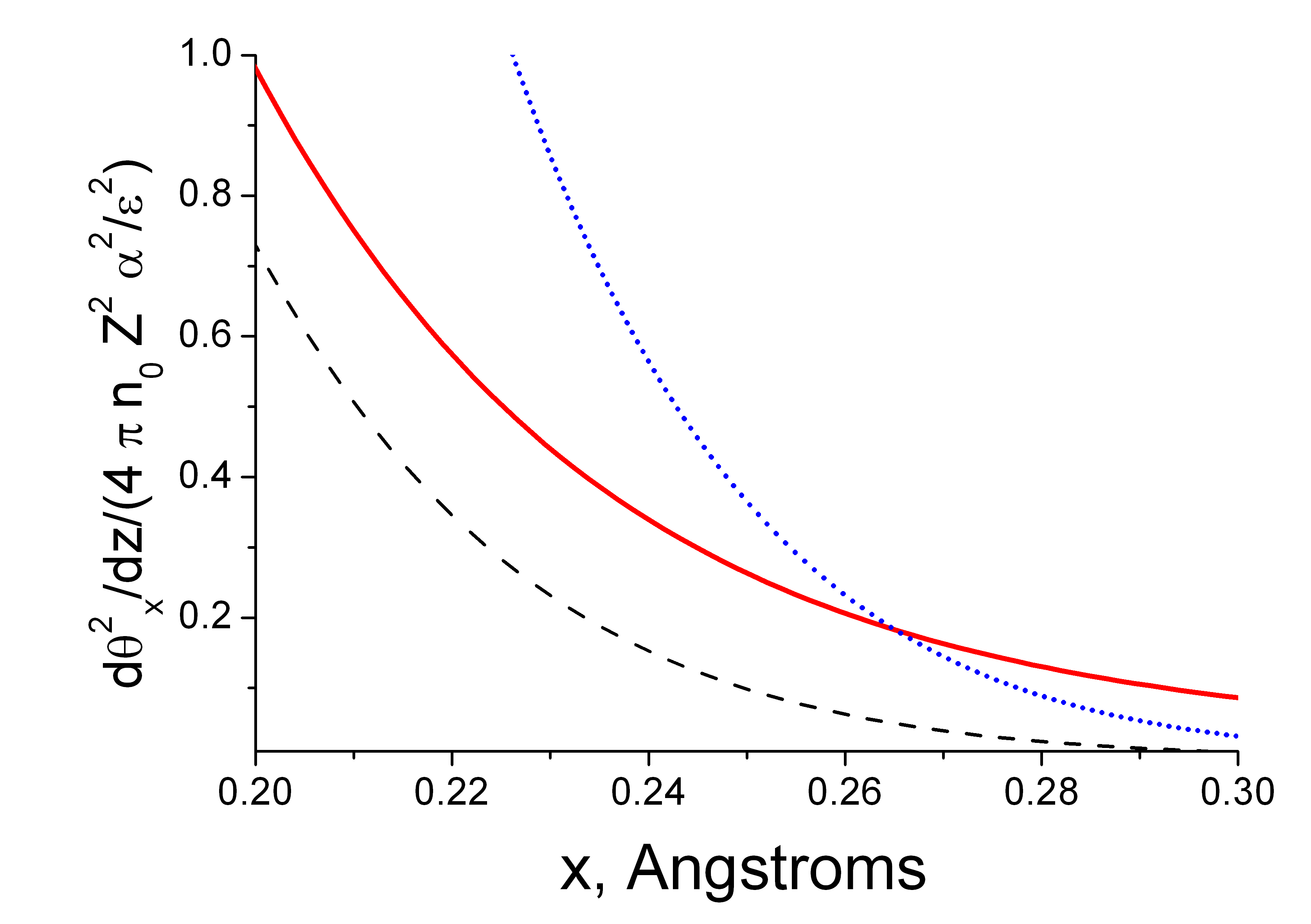}} \\
\caption{The x-coordinate dependence of the local mean square
projected scattering angles per unit length: solid red line --
determined by Eq. (30), dashed black -- by the same with $\delta
\ln(x) = 0$ and dotted blue -- by Eq. (A2)). All the values are
normalized to $4 \pi Z^2 n_0 \alpha^2 /\varepsilon^2$. The same
data are depicted on the left and on the right, in logarithmic and
linear scales, respectively}
\end{center}
\end{figure}

At $\delta \ln(x) > 0 $, on the opposite, the Coulomb scattering
is sampled for any $q > 0$, however, to take into consideration
the scattering on individual phonons, described by the large
positive logarithm modification (29) from Fig. 4, a random
scattering, characterized by the mean square angle
\begin {equation} 
\left\langle  \frac{d \theta^2_x (x, q_2)}{d z}\right\rangle =
\frac{4 Z^2 \alpha^2 n(x)}{p^2 v^2} \; \delta \! \ln (x),
\end{equation}
should be additionally sampled. Fig. 4 demonstrates that the
latter can reach drastic relative values, reflecting the distant
one-phonon excitation predominance over the local scattering.

To compare the roles of different scattering mechanisms, the
corresponding mean square angles are compared in Fig. 5 in the
case of (110) Si planes and positron energy of 1.5 GeV,
corresponding to the "optimal" crystal undulator construction,
suggested in \cite{tik2}. To reveal the role of the Coulomb
scattering at local density, the solid curve, evaluated using Eq.
(30), is compared with the dashed one, evaluated using the same
with $\delta \ln(x)$ put to zero. While the latter curve
approaches the former one at the considerably nuclear densities,
at $x \sim 2 u_1$ it becomes inferior, when the scattering on
individual phonons dominates. Comparing further the solid curve
with the dotted one, built according to Eq. (A2) to represent the
predictions of the local scattering approximation \cite{tar,sca3},
one concludes that the latter both overestimates the scattering
intensity at the high, and underestimates it at the low nuclear
densities. The difference in prediction of Eqs. (30) and (A2)
would be even deeper if the overestimated Coulomb logarithm of the
latter has not been compensated by the multiplication by the
vanishing nuclear density. Note that the particular value of $q_2$
in Eq. (30), which was fixed by the condition $q_2^2/2 \varepsilon
= U(2u_1)$, has a marginal effect on the revealed predictions.

It should be clarified, that the described sampling receipts do
not precisely reflect the real physical peculiarities of the
scattering process, representing only the technical means to
incorporate the thoroughly evaluated mean square incoherent
scattering angles into the classical trajectory simualtions.

In general, the overestimated growth of incoherent scattering
intensity (the dotted blue curve in Fig. 5) supports the
assumption \cite{bir} of the instant "turning on" of the nuclear
scattering at some distance $x \sim 1/\kappa$ from the plane.
Being adopted, in fact, from the nonrelativistic ion channeling
theory, this approach loses its applicability at multi-Gev
energies. The demonstrated above slowdown of the decrease of
incoherent scattering intensity with transverse coordinate (see
the solid red curve in Fig. 5), on the opposite, strengthens the
picture of the nuclear dechanneling process widely distributed
both in longitudinal and transverse directions \cite {tik6}, being
poorly treated by the single dechanneling length approach.

\subsection{The lack of a consistent electron scattering theory}

The process of incoherent particle scattering by the electrons of
medium atoms, called shortly the electron scattering, gives a
roughly $Z$-times smaller contribution to the mean square Coulomb
scattering angle than the nuclear one, being often treated
approximately \cite{mol,bet}. However, the electron scattering can
take over the leading role in the low nuclear density regions of
the stable positively charged particle channeling, making
necessary its quantitative consideration in crystals.

Contrary to the multiple scattering, particle collisions with
electrons are essential for the ionization energy loss process,
the theory of which is well developed accordingly
\cite{bet,LL3,LL4,LL8,jac}. The latter reveals that relativistic
particles efficiently transfer their energy to electrons at impact
parameters far exceeding the inter-atomic distance in condensed
medium, making the local consideration of the ionization losses in
crystals even less appropriate, than for the nuclear scattering.

Though the treatment of the ionization energy losses in the
inhomogeneous electron distribution of atomic strings and planes
is already known \cite{esb,esb2,bur}, its application to the
process of electron multiple scattering is not straightforward by
the reason that particle deflection is related solely with the
transverse momentum transfer, while the ionization losses are with
both transverse and longitudinal ones. Giving no contribution to
the multiple scattering, the latter dominates at relativistic
ionization energies. However, since the transverse channeling
motion becomes relativistic \cite{bar,bar2,akh,bai,kim} at the
particle energy of $m^2/2 V_0 \sim 0.1-10$ GeV, the latter case
becomes quite common in the TeV-energy region, making the seminal
ionization loss formula \cite{pdg} inapplicable for the mean
square scattering angle evaluation and revealing thus the need in
its alternative.

Longitudinal momentum component also dominates near the threshold
of the atom excitation or ionization, which determines the minimal
value of the former, being essential for both the Coulomb
logarithm and density effect reconsideration. Finally, the
electron multiple scattering theory should be enriched by the
temperature depending Debye-Waller factor, both the partial and
full atomic form factors \cite{fer,fer2} as well as the incoherent
scattering function \cite{fan, hub}.

\section{Conclusions}
The theory, which describes the incoherent particle scattering by
atomic planes with neither arbitrary assumptions nor
phenomenological parameter introduction has been developed. The
obtained formula for the mean square scattering angle of
incoherent scattering proves to be in agreement with the
alternative approach \cite{tik7}, developed for the axial case.
Both of them predict the modest incoherent scattering suppression
in the dense nuclear regions as well as the inapplicability of the
local scattering treatment far from the latter. All the quantum
features of incoherent scattering are accumulated in the Coulomb
logarithm modification (29). The latter secures the minor
amendments of the Coulomb scattering sampling, which take into
consideration the influence of the crystal plane atom distribution
heterogeneity on the incoherent scattering of relativistic
particles moving along classical trajectories. Though the present
theory has been developed for the case of channeling motion, we
envisage its wider application, demonstrated by the agreement both
with the uniform particle flux limit (31) \cite{ter2,maz2} and the
axial case \cite{tik7}. Similarly to latter, the developed
approach makes it possible to refine the numerical treatment of
any experiment on high energy particle scattering and radiation in
the fields of crystal planes ever conducted or planned.

Financial support by the European Commission through the N-Light
Project, Grant Agreement number: 872196, is gratefully
acknowledged.

\section{Appendix A}
The developed incoherent scattering consideration should be
integrated into the established Coulomb scattering theory. The
most acknowledged \cite{Gea} version of the latter proceeds from
the simple screened Coulomb cross section
$$ 
 \frac{d^2 \sigma_C(q)}{dq^2} = \frac{4 Z^2
\alpha^2}{v^2 (q^2+\kappa^2)^2}, \eqno (A1)
$$
leading to the widely used formula
$$ 
\frac{d \theta_{space}^2}{dz} =  2\frac{d \theta_0^2}{dz} = n
\int_0^{\vartheta_{max} p } \frac{q^2}{ p^2} \frac{d^2
\sigma_C(q)}{dq^2} dq^2  = \left( \frac{21 MeV}{v p} \right)^2
L_{rad}^{-1} \simeq \frac{8 \pi Z^2 \alpha^2 n}{p^2 v^2} \ln
\frac{\vartheta_{max}}{\vartheta_{min}} \eqno (A2)
$$
for both the mean square nonprojected (space) $ d_{space}^2/dz$
and projected (plane) $d \theta_0^2/dz$ scattering angles (we use
both the notations and terminology of \cite{pdg}) in amorphous
media with atomic number $Z$ and nuclear density $n$,
$\vartheta_{min} = \kappa/p$ and $\vartheta_{max}$ are
respectively the typical atom and maximal nuclear scattering
angles. Since the atomic plane consideration assumes the azimuthal
symmetry violation, we consider mainly the mean square projected
(plane) angles in both the channeling $xz$ (27), (28), (30), (B3)
and the normal to it $yz$ (B4) planes, both of which should be
compared with the same in amorphous media $d \theta_0^2/dz$,
being, according to Eq. (A2), twice as less as the nonprojected
(space) one. In particular, we will use the mean square projected
scattering angle for both the limited, $0 < q < q_2$,
$$ 
\frac{d \theta_0^2(q<q_2)}{dz} =  \frac{1}{2} n \int_0^{q_2}
\frac{q^2}{ p^2} \frac{d^2 \sigma_C(q)}{dq^2} dq^2 = \frac{4 \pi
Z^2 \alpha^2 n}{v^2 p^2} \left[ \ln \left( 1 +
\frac{q^2_2}{\kappa^2}\right)^{1/2} - \frac{q^2_2}{2
(q^2_2+\kappa^2)}\right] \eqno (A3)
$$
and asymptotic, $q_1 \gg \kappa$, $q_2 \gg q_1$,
$$ 
\frac{d \theta_0^2(q_1<q<q_2)}{dz} =  \frac{1}{2} n
\int_{q_1}^{q_2} \frac{q^2}{p^2} \frac{d^2 \sigma_C(q)}{dq^2} dq^2
\simeq \frac{4 \pi  Z^2 \alpha^2 n}{v^2 p^2} \ln \frac{q_2}{q_1}
\eqno (A4)
$$
momentum transfers, the latter of which corresponds to the high
momentum transfer limit
$$ 
\displaystyle  {\left\langle  \frac{d \theta^2_x (x)}{d
z}\right\rangle \simeq \frac{4 Z^2 \alpha^2 n_0 d}{\pi p^2 v^2 }
\int_{q_1}^{q_2} \int_0^{2 \pi} \int_{-\infty}^\infty \frac{q^2_x
\exp (2 i k x-2 k^2 u^2_1)}{(q^2 + \kappa^2)^2}  dk \: d\varphi \:
q dq } \eqno (A5)
$$
of Eq. (27) at local nuclear density $n = n_n(x)$, given by Eq.
(4), as is readily seen from the integral
$$ 
\displaystyle { \int_{-\infty}^\infty \exp (2 i k x -2 k^2 u^2_1)
dk = \frac{\sqrt{\pi}}{\sqrt{2}\,u_1}  \exp \left(-\frac{x^2}{2
u_1^2} \right) = \frac{\pi n_n(x)}{n_0 d} } \eqno (A6)
$$
and will be used to extract the effective local mean square
projected angle per unit length (27) for arbitrary momentum
transfers.

\section{Appendix B}
The central result (27) is reproduced below using the Wigner
function
$$ 
W_{\bm\rho_n}(\bm\rho, \bm q)= -\displaystyle{\frac{8 Z
\alpha}{\pi v}} \displaystyle{\left [\frac{\cos 2 \bm q (\bm\rho -
\bm\rho_n) - \cos (2 \bm q \bm\rho) \exp ( - 2 q^2 u_1^2)}{4 q^2 +
\kappa^2} \right]} $$
$$ 
 + \displaystyle{ \frac{4 Z^2 \alpha^2
}{\pi^2 v^2} \int \exp(2 i \textbf{k} \bm\rho) \left\{ \frac{\exp
[- i (\bm q + \textbf{k} ) \bm\rho_n] -  \exp [- (\bm q +
\textbf{k} )^2 u^2_1/2 ]}{(\bm q + \textbf{k})^2+\kappa^2}
\right\} }$$
$$ 
\displaystyle{ \times\left\{\frac{\exp [i (\bm q - \textbf{k})
\bm\rho_n] -  \exp [- (\bm q - \textbf{k} )^2 u^2_1/2 ]}{(\bm q -
\textbf{k})^2+\kappa^2} \right\} } d^2 k \eqno (B1) $$
$$ 
- \displaystyle{ \frac{4 Z^2 \alpha^2 }{\pi^2 v^2} \exp(2 i \bm q
\bm\rho) \int \left\{ \frac{\exp [- i (\bm q + \textbf{k})
\bm\rho_n] - \exp [- (\bm q + \textbf{k} )^2 u^2_1/2 ]}{(\bm q +
\textbf{k})^2+\kappa^2} \right\} }$$
$$  
\displaystyle{ \times\left\{\frac{\exp [-i(\bm q - \textbf{k})
\bm\rho_n] -  \exp [- (\bm q - \textbf{k} )^2 u^2_1/2 ]}{(\bm q -
\textbf{k})^2+\kappa^2} \right\} } d^2 k,
$$
obtained by an alternative method in the axial case \cite{tik7}.
The mean square angles of incoherent scattering in the planes $xz$
and $yz$
$$  
\left\langle  \frac{ \theta^2_i (\bm\rho)}{d z}\right\rangle  =
\int\int\int\int \frac{q_i^2}{p^2} W_{\bm\rho_n}(\bm\rho, \bm q)
n_n(\rho_n,0) d^2q d^2 \rho_n = \frac{4 Z^2 \alpha^2 }{\pi^2 v^2
p^2 d_{at}}
$$
$$ 
\times\int\int\int\int \frac{\exp [- 2 k^2 u_1^2] - \exp [- (k^2 +
q^2) u^2_1]}{([\bm q + \textbf{k})^2+\kappa^2] [(\bm q -
\textbf{k})^2+\kappa^2]} (q_i^2 - k_i^2) \exp(2 i \textbf{k}
\bm\rho) d^2 k d^2q \eqno (B2)
$$
are obtained by Eq. (B1) multiplication by $q_i^2$, $i=x,y$,
integration over $\textbf{q}$ and averaging over the string atom
nuclei distribution, following from Eq. (14) at $\psi = 0$. Both
the integrand symmetry and variable notation interchange of
$\textbf{q} \leftrightarrow \textbf{k}$ have been used in the
above transformation as well. To reduce Eq. (B2) to the planar
case, it is integrated further over $y= \psi z$, resulting in the
Dirac delta function $\delta(2 k_y)$, absorbed by the integration
over $k_y$, yielding both the mean square scattering projected
(plane) angle per unit length in both the $xz$
$$ 
\left\langle  \frac{ \theta_x^2 (x)}{d z}\right\rangle
 =  \int \left\langle  \frac{ \theta_x^2
(\bm\rho)}{d z}\right\rangle \frac{dy}{d_{st}} = \frac{4 Z^2
\alpha^2 n_0 d}{\pi p^2 v^2}
$$
$$  
\times  \int\int\int \frac{\exp (- 2 k^2 u_1^2) - \exp [- (k^2 +
q^2) u^2_1]}{[(q_x+ k)^2+q_y^2+\kappa^2]\: [(q_x -
k)^2+q_y^2+\kappa^2]}(q_x^2 - k^2) \exp(2 i k x)\, d k\, d^2q
\eqno (B3 \equiv 27)
$$
and $yz$ plane
$$
\left\langle  \frac{ \theta_y^2 (x)}{d z}\right\rangle
 =  \int \left\langle  \frac{ \theta_y^2
(\bm\rho)}{d z}\right\rangle \frac{dy}{d_{st}} = \frac{4 Z^2
\alpha^2 n_0 d}{\pi p^2 v^2}
$$
$$
\times  \int\int\int \frac{\exp (- 2 k^2 u_1^2) - \exp [- (k^2 +
q^2) u^2_1]}{[(q_x+ k)^2+q_y^2+\kappa^2]\: [(q_x -
k)^2+q_y^2+\kappa^2]} q_y^2 \exp(2 i k x)\, d k\, d^2q, \eqno (B4)
$$
the latter of which can alternatively be obtained using
$\hat{H}_0\varphi = \hat{p^2_y} / 2 \varepsilon $ instead of Eq.
(3).

\section{Appendix C}
To demonstrate the asymptotic behavior of Eq. (27) at large $x$,
let us represent its first integral in the form
$$ 
\displaystyle  { \int_{-\infty}^\infty  \int_{-\infty}^\infty
\int_{-\infty}^\infty \frac{ \exp [-(q^2_x + k^2)u^2_1 + 2 i k
x]\times (q^2_x -k^2) }{[(q_x + k)^2 + q_y^2 + \kappa^2] [ (q_x -
k)^2 + q_y^2 + \kappa^2]} \sum_{n=0}^\infty \frac{1}{n!}[(q^2_x
-k^2) u^2_1]^n  dk \: dq_x \: dq_y.} \eqno (C1)
$$ 
All the integrals both over $k$ and $q_x$ are connected with the
one
$$ 
\int_{-\infty}^\infty \frac{ \exp(- k^2 u_1^2/2 + i k x)
}{k^2+q^2_y+\kappa^2} dk  = \exp [(q^2_y + \kappa^2) u^2_1/2 ]
\left( \int_{-\infty}^\infty \frac{ \exp[- \alpha (k^2 + q^2_y +
\kappa^2) u_1^2/2 + i k x] }{k^2+q^2_y+\kappa^2} dk
\right)_{\alpha = 1}
$$
$$ 
=\frac{\pi}{\sqrt{q_y^2 + \kappa^2}} \exp {[(q^2_y + \kappa^2)
u^2_1/2 -\sqrt{q_y^2 + \kappa^2}\, x] }
$$
$$- \sqrt{2 \pi} u_1 \exp [(q^2_y + \kappa^2) u^2_1/2 ]
\int_0^1\exp \left[ - \frac{1}{2} \beta^2 (q^2_y + \kappa^2) u^2_1
- \frac{x^2}{2 \beta^2 u^2_1} \right] d \beta
$$
$$ 
\simeq \frac{\pi}{\sqrt{q_y^2 + \kappa^2}} \exp \left[(q^2_y +
\kappa^2) u^2_1/2-\sqrt{q_y^2 + \kappa^2} \, x\right]. \eqno (C2)
$$
In order to separate the local contribution of the nuclear density
above, a first order differential equation has been introduced by
means of taking a derivative over the fictitious parameter
$\alpha$. Eq. (C2) can be readily applied to approximate the
integral
$$ 
 \displaystyle {
 \int_{-\infty}^\infty \frac{ sin(k
x ) \exp(-k^2 u_1^2/2) }{k^2+q^2_y+\kappa^2} k dk  = \pi \exp
[(q^2_y + \kappa^2) u^2_1/2 -\sqrt{q_y^2 + \kappa^2}\, x] }
$$
$$ 
 - \sqrt{2 \pi} \frac{x}{u_1} \exp [(q^2_y + \kappa^2) u^2_1/2 ] \int_0^1\exp \left[ - \frac{1}{2}
\beta^2 (q^2_y + \kappa^2) u^2_1 - \frac{x^2}{2 \beta^2 u^2_1}
\right] \frac{d \beta }{\beta^2}
$$
$$
\simeq \pi \exp \left[(q^2_y + \kappa^2) u^2_1/2-\sqrt{q_y^2 +
\kappa^2} \, x\right],\eqno (C3)
$$
corresponding both to the $n=0$ term of the expansion (C1) and the
second integral of Eq. (27), as well as the one of
$$ 
\displaystyle {
 \int_{-\infty}^\infty \frac{ k^2 cos(k
x ) \exp(-k^2 u_1^2/2) }{k^2+q^2_y+\kappa^2} dk = \frac{\sqrt{2
\pi}}{u_1} exp \left(-\frac{x^2 }{2 u^2_1}\right) }
$$
$$ 
\displaystyle { -  (q^2_y+\kappa^2) \int_{-\infty}^\infty \frac{
cos(k x ) \exp(-k^2 u_1^2/2) }{k^2+q^2_y+\kappa^2} dk }
$$
$$\simeq
-\pi \sqrt{q_y^2 + \kappa^2} \exp \left[(q^2_y + \kappa^2)
u^2_1/2-\sqrt{q_y^2 + \kappa^2} \, x\right],   \eqno (C4)
$$
which represents itself the $n=1$ term of the expansion (C1). The
above integrals demonstrate that the Gaussian local nuclear
density contributions to Eq. (27) are negligible at $x \gg u_1$.
Reducing them further to the MacDonald functions $K_0$, $K_1$, one
arrives to the leading term of the Eq. (27) expansion in series in
small parameter $u_1 \kappa$
$$ 
\displaystyle  {\left\langle  \frac{d \theta^2_x (x,q_2,q_1)}{d
z}\right\rangle \simeq \frac{4 \pi Z^2 \alpha^2 n_0 d }{p^2 v^2 }
\kappa (u_1 \kappa)^2 \left [ K_1(2 \kappa x )+ \frac{1}{ \kappa
x} K_0(2 \kappa x ) + \frac{1}{ \kappa^2 x^2} K_1(2 \kappa x )
\right ].} \eqno (C5)
$$ %
Note that the leading contributions to both the first and the
second integrals of Eq. (27) greatly exceed (C5), however cancel
each other. Since MacDonald functions exponentially decrease as
$$ 
K_0(2 \kappa x ), K_1(2 \kappa x ) \sim \frac{1}{2}
\sqrt{\frac{\pi}{\kappa x}} \exp{(-2 \kappa x) } \eqno (C6)
$$
at $ x
> 1/\kappa$, Eq. (C5) visualizes the expected non-gaussian nonlocal behavior of the mean
square scattering angle (27) at large distances from the atomic
plane, corresponding both to the explosive logarithm growth,
depicted in Fig. 4, and the decelerated local mean square angle
decrease, illustrated in Fig. 5. As in the axial case \cite{lin},
the factor $(\kappa u_1)^2$ appears in Eq. (C5) after the
averaging of the square of the averaged atom field $2 \pi Z \alpha
\, exp(-\kappa x)$ thermal perturbation.

Though the higher $n = 2, 3,..$ terms of the asymptotic expansion
(C1) can be readily evaluated, the one-dimensional integral
representation of the first three-dimensional integral of Eq. (27)
proves to be more practical. In addition, the removal (C3) of the
oscillating integrand simplifies the evaluation of Eqs. (27)-(29).

\end{document}